\documentclass[aps,prst,preprint,groupedaddress]{revtex4-1}

\usepackage{mathrsfs,amssymb,graphics,subfigure,threeparttable}
\usepackage{grap hicx}
\usepackage{amssymb}
\usepackage{enumerate}
\usepackage{amsmath}
\usepackage{fancyhdr}
\usepackage{bm}
\usepackage[squaren]{SIunits}

\begin{document}


\title{Low emittance electron beam generation from a laser wakefield accelerator using two laser pulses with different wavelengths}


\author{X. L. Xu}
\affiliation{Tsinghua University, Beijing 100084, China}
\author{Y. P. Wu}
\affiliation{Tsinghua University, Beijing 100084, China}
\author{C. J. Zhang}
\affiliation{Tsinghua University, Beijing 100084, China}
\author{F. Li}
\affiliation{Tsinghua University, Beijing 100084, China}
\author{Y. Wan}
\affiliation{Tsinghua University, Beijing 100084, China}
\author{J. F.  Hua}
\affiliation{Tsinghua University, Beijing 100084, China}
\author{C.-H. Pai}
\affiliation{Tsinghua University, Beijing 100084, China}
\author{W. Lu}
\email[]{weilu@tsinghua.edu.cn}
\affiliation{Tsinghua University, Beijing 100084, China}
\affiliation{University of California Los Angeles, LA, CA 90095, USA}
\author{P. Yu}
\affiliation{University of California Los Angeles, LA, CA 90095, USA}
\author{C. Joshi}
\affiliation{University of California Los Angeles, LA, CA 90095, USA}
\author{W.B. Mori}
\affiliation{University of California Los Angeles, LA, CA 90095, USA}


\date{\today}

\begin{abstract}	
Ionization injection triggered by short wavelength laser pulses inside a nonlinear wakefield driven by a longer wavelength laser is examined via multi-dimensional particle-in-cell simulations. We find that very bright electron beams can be generated through this two-color scheme in either collinear propagating or transverse colliding geometry. For a fixed laser intensity $I$, lasers with longer/shorter wavelength $\lambda$ have larger/smaller ponderomotive potential ($\propto I \lambda^2$). The two color scheme utilizes this property to separate the injection process from the wakefield excitation process. Very strong wakes can be generated at relatively low laser intensities by using a longer wavelength laser driver (e.g. a $10~\micro\meter$ CO$_2$ laser) due to its very large ponderomotive potential. On the other hand, short wavelength laser can produce electrons with very small residual momenta ($p_\perp\sim a_0\sim \sqrt{I}\lambda$) inside the wake, leading to electron beams with very small normalized emittances (tens of $\nano\meter$). Using particle-in-cell simulations we show that a $\sim10~\femto\second$ electron beam with $\sim4~\pico\coulomb$ of charge and a normalized emittance of $\sim 50~\nano\meter$ can be generated by combining a 10 $\micro\meter $ driving laser with a 400 $\nano\meter$ injection laser, which is an improvement of more than one order of magnitude compared to the typical results obtained when a single wavelength laser used for both the wake formation and ionization injection. 
\end{abstract}

\pacs{}

\maketitle

\section{Introduction}
Recently, significant progress has been made in the field of plasma-based acceleration \cite{RevModPhys.81.1229}. In the laser driver case (laser wakefield acceleration LWFA), $\giga\electronvolt$ level energy gain has been shown in recent experiments \cite{leemans2006gev, wang2013quasi, PhysRevLett.111.165002, PhysRevLett.105.105003}. In the beam driver case (plasma wakefield acceleration PWFA), energy gains up to $42~\giga\electronvolt$ have been demonstrated \cite{PhysRevLett.93.014802, PhysRevLett.95.054802, blumenfeld2007energy}. These experiments have shown the ability of relativistic plasma wakes to accelerate electrons at ultra-high gradients over a significant distance. A major focus of current research in this field is to significantly improve the quality of the electron beams generated by plasma accelerators. This is because electron beams with smaller energy spreads and transverse emittances than those currently obtained from plasma accelerators are needed for future applications of this technology. These applications include free-electron lasers \cite{Barletta201069} and linear colliders \cite{leemans2009laser}. Towards this goal, many controlled injection schemes \cite{PhysRevLett.102.065001}\cite{PhysRevLett.100.215004} have been proposed and demonstrated in the past few years. Among these schemes, the ionization-based injection method \cite{PhysRevLett.98.084801, PhysRevLett.104.025003, PhysRevLett.107.045001} has attracted much attention due to its simplicity and flexibility. In ionization injection in a LWFA driven by a single laser pulse, mixed gases (e.g., He and N$_2$) with different ionization potentials (IP) are used to provide electrons both for forming the wake and injection. The wake is formed by ionization of the low IP electrons during the rise time of the laser pulse. The higher IP electrons of a high-Z atom (e.g., N$_2$) are released near the peak of the laser intensity inside a fully formed wake and then slip back to the tail of the wake. If a trapping condition is satisfied \cite{PhysRevLett.98.084801}\cite{PhysRevLett.104.025003} then electrons gain enough energy as they slip backwards so that they eventually move forward with speeds in excess of the wake's phase velocity. This scheme has an obvious drawback for generating very low beam emittances: to drive a large wake so that trapping can occur \cite{PhysRevLett.104.025003}, a laser pulse with a normalized vector potential $a_0$ greater than $1$ is needed, which leads to a large residual momentum on the order of $\sim a_0$ for the injected electrons along the laser polarization direction even in one-dimensional (1D). At the same time, the radius of the ionized region of injected electrons, which is connected with the radius of the driving laser, is on the order of a few laser wavelengths or a few $\micro\meter$. The transverse ponderomotive force for such a radius can also lead to residual momentum in both transverse directions. These residual momentum and initial radius lead to a normalized emittance on the order of 1 $\micro\meter$ in the both transverse directions. 

To obtain a much smaller emittance through ionization injection, a possible strategy is to separate the wake excitation process and the injection process so that the injection laser can have a much lower vector potential and much smaller spot size. This strategy has been demonstrated recently in two schemes utilizing an electron beam driver to excite the wake \cite{PhysRevLett.108.035001}\cite{PhysRevLett.111.015003}. Due to the much lower electric field of the electron beam driver, ionization of low IP atoms (e.g., H or Li) are used to excite the wake while the modest IP helium atoms can be used to provide the injected electrons, which requires a much lower laser intensity. For example, the intensity threshold to ionize the first electron of He is $\sim10^{15}~\watt\per\centi\meter^2$ which corresponds to $a_0\sim0.02$ for an 800 $\nano\meter$ wavelength laser. On the other hand for a single laser driver, the intensity threshold to ionize N$^{5+}$ requires an intensity of $\sim10^{19}~\watt\per\centi\meter^2$ or $a_0\sim2$ for an 800 $\nano\meter$ laser. Also, laser spot sizes much smaller than the transverse size of the wake can be used to reduce the radius of the ionized region. The use of lower $a_0$ and smaller spot sizes lead to much reduced normalized emittances of $\sim 30~\nano\meter$ when using a single co-propagating laser pulse in a wake driven by a particle beam \cite{PhysRevLett.108.035001}. Furthermore, if the injection time can be limited through the use of two transversely colliding laser pulses then emittances less than 10 $\nano\meter$ can potentially be achieved \cite{PhysRevLett.111.015003}. In addition, the dependence on the evolution of the emittance of particles based on the initial spot size and ionization time has recently been quantified \cite{PhysRevLett.112.035003}.

In this paper, we explore through multi-dimensional particle-in-cell (PIC) simulations a natural extension of separating the wake excitation from the ionization injection. In this all optical concept, two laser pulses with different wavelengths --- a long wavelength ($\sim 10~\micro\meter$) laser pulse with a large ponderomotive potential but small intensity is used to drive the wake and a short wavelength ($\sim 400~\nano\meter$) laser pulse with a small ponderomotive potential but a large intensity is used to inject electrons into the wake. Since a long wavelength laser pulse has a large ponderomotive potential, it still produces a large amplitude wake at a relatively low intensity. This is analogous to the beam-driven wake where a much lower beam electric field can be used to drive a strong wake. The use of two lasers with different wavelengths was recently studied using 1D PIC simulations and theory \cite{yu2013low}. However, due to the intrinsic multi-dimensionality of the problem, multi-dimensional PIC simulations are needed to obtain quantitative results. For example, the emittance of the beam is determined by the spot size over which the electrons are ionized and the transverse ponderomotive force from the second laser. We also explore using transversely propagating lasers to trigger the ionization which can not studied in 1D. Our PIC simulation results show that it is possible to generate electron beams containing $\sim 4~\pico\coulomb$ of charge with small normalized emittances ($\sim 50~\nano\meter$),  more than one order of magnitude smaller than that obtained in ionization injection using a single 800 $\nano\meter$ wavelength laser. 

Throughout this paper the injection of electrons into the plasma wake is due to ionization injection, rather than by self injection in a preformed plasma in a very nonlinear wakefield \cite{PhysRevSTAB.10.061301}\cite{PhysRevLett.103.215006}. The simulations were performed using the particle-in-cell (PIC) code OSIRIS \cite{fonseca2002high}. The parameters for all simulations for which results are presented are provided in Table \ref{tab: simulation paras}. In section II, a simulation (A) of a single 800 $\nano\meter$ laser pulse, propagating in a mixture of He and N$_2$ gases, that both excites the wake and injects through ionization of the K-shell of the N$_2$ is presented as a reference case. It is assumed that the He gas is fully ionized. Another simulation (B) of a single 800 $\nano\meter$ laser pulse with lower intensity is also presented in section II. Then in section III and IV, results from simulations using two different color laser pulses to trigger injection using co- and transversely propagating lasers are presented (C, D, E, F). Lastly, a summary of the findings is given in section V.

\begin{table}
\begin{threeparttable}
\caption{\label{tab: simulation paras} Simulation parameters of the presented results}
\begin{ruledtabular}
\begin{tabular}{ccccccc}
\multicolumn{1}{c}{Simulation parameters} & A & B & C & D & E & F\\
\hline
Box size [$k_0^{-1}$] & $[450, 500]$ & $[350, 500]$ & $[250, 500]$ & $[300, 400]$ & $[400, 400]$ & $[400, 400]$\\
Number of cells & $[4500, 1600]$ & $[1750, 1000]$ & $[1250, 2500]$ &$[9000, 800]$ & $[50000, 800]$ & $[2000, 20000]$\\
Particles per cell: plasma electrons & $[1, 1]$ & $[1, 1]$ & $[1, 1]$ & $[1, 1]$ & $[1, 1]$ & $[1, 1]$\\
Particles per cell: ions & $[4, 4]$ & $[2, 2]$ & $[4, 4]$ & $[4, 4]$ & $[2, 2]$ & $[2, 2]$\\
Time step $[\omega_0^{-1}]$ & 0.05 & 0.1 & 0.1 & 0.033 & 4e-3 & 1.99e-2\\
Preformed plasma density $[10^{17}~\centi\meter^{-3}]$ & 21 & 21 & 21 & 17 & 0.12 & 0.12 \\
Ionized gas density $[10^{17}~\centi\meter^{-3}]$ & 0.21 & 0.21 & 0.21 & 0.34 & 0.024 & 0.024 \\
Laser wavelengths [\micro\meter]: $\lambda_0$, $\lambda_1$ & 0.8 & 0.8 & 0.8, 0.8 & 0.8, 0.08 & 10, 0.4 & 10, 0.4 \\
Laser amplitudes: $a_0$, $a_1$ & 2.0 & 1.4 & 1.2, 2.0 & 1.2, 0.25 & 1.4, 0.09 & 1.4, 0.06 \\
Laser pulse lengths [$\femto\second$]: $\tau_0$, $\tau_1$ & 32 & 51 & 32, 12 & 18, 7 &  352, 44 & 352, 13 \\
Laser pulse delay [\femto\second] & none & none & 40 & 30 & 584 & 245 \footnote{The delay between the collision position and the center of the drive laser in the transverse injection case.} \\
\end{tabular}
\end{ruledtabular}
\end{threeparttable}
\end{table}

\section{Factors affecting beam emittance in ionization injection and wake excitation using a single laser pulse}
\begin{figure}[bp]
\includegraphics[width=1.0\textwidth]{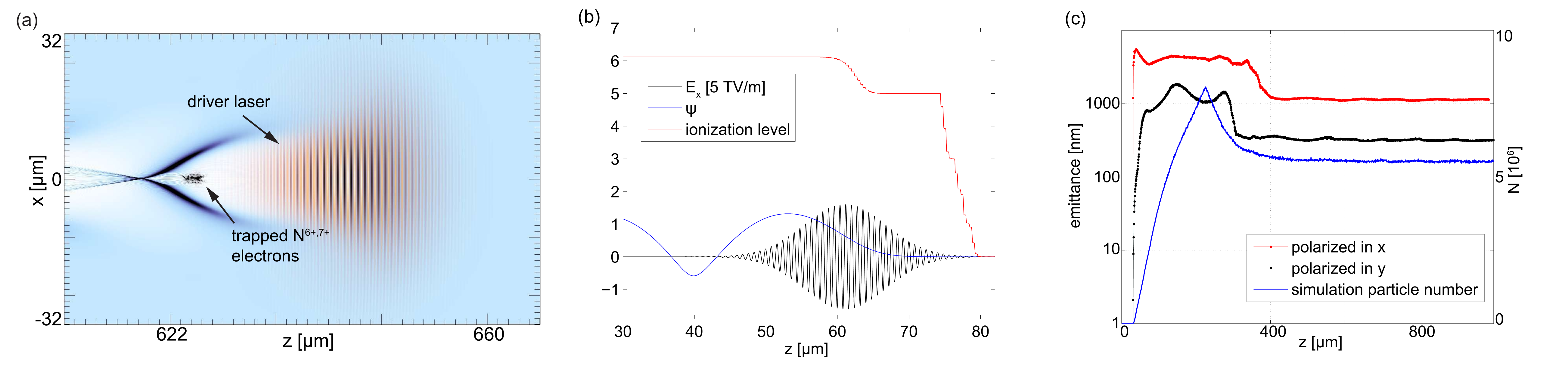}
\caption{\label{fig: simulation A} 2D OSIRIS simulation (A) of wake excitation and ionization injection driven by a single laser pulse ($n_e=2.1\times 10^{18}~\centi\meter^{-3}$). A 800 $\nano\meter$ laser pulse with $w_0=13~\micro\meter$ and $a_0=2$ propagates to the right in a mix of pre-ionized plasma  and $N_2$ ($n_N = 0.01n_e$). The laser pulse  has a longitudinal Gaussian profile with $\tau_{FWHM}= 32~\femto\second$.  The nitrogen distribution is limited to the first 190 $\micro\meter$. (a) Snapshot of the charge density distribution of the wake electrons, the K-shell electrons of nitrogen, and the electric field in $x$ direction.  (b) The laser electric field $E_x$, the normalized pseudo potential $\psi$ of the plasma wake and the ionization state of nitrogen atoms on axis. (c) The emittance evolution of the beam along (red) and perpendicular to the polarization direction (black), and the evolution of the particle number in the simulation (blue). The saturated emittance are 1200 $\nano\meter (\hat{x})$  and 350 $\nano\meter (\hat{y})$ for the laser polarized in and out of the simulation plane respectively. Due to the particle loss as shown in the blue line, the saturated emittance is smaller than the thermal emittance which contains all the ionized particles. In this simulation the injection distance is controlled by the longitudinal nitrogen distribution. The injected charge is about 40 $\pico\coulomb$ by assuming a cylindrical distribution around $z$-axis. }
\end{figure}

To understand what factors limit the ultimate beam emittance achieved when the wake is both excited and the charge is injected via ionization injection by a single laser pulse, we follow the emittance evolution process of the injected electrons by particle tracking in PIC simulations. We present results from simulation A in Table \ref{tab: simulation paras}. Fig. \ref{fig: simulation A}(a) shows a typical scenario for ionization injection using a single laser pulse, where an intense short pulse laser propagates through a mixture of helium and nitrogen gases. The simulation is initialized with the two electrons from helium atoms being fully ionized and the first five L-shell electrons from nitrogen atoms are stripped off at the very leading edge of the laser pulse and are pushed out by the ponderomotive force of the laser, thereby forming the nonlinear wake. The two K-shell electrons of nitrogen atom have very high ionization potentials (552 and 667 $\electronvolt$ respectively), therefore they can only be freed in the middle of the wake near the peak of the laser intensity  (e.g., $10^{19}~\watt\per\centi\meter^2$ or $a_0 \sim 2$ for 800 $\nano\meter$ laser wavelength). This is evident in the plots of the ionization levels, the laser electric field, and the wake potential shown in Fig. \ref{fig: simulation A}(b). After ionization, these K-shell electrons start to gain energy in the wake while still slipping backwards relative to the wake until their longitudinal velocity approaches the phase velocity of the wake. After this point, they are trapped in the wake and then keep gaining energy until they outrun the wake by dephasing. This trapping process may be stopped by the nonlinear wake evolution caused by the driver evolution or the beam loading effect after enough electrons become trapped \cite{PhysRevLett.101.145002}. In principle one can control the charge, the energy spread and the emittance of the injected electrons by modifying the density profile and the distance occupied by nitrogen gas, and/or by varying the nitrogen concentration. However, here we consider controlling the beam parameters through the laser parameters.

Throughout this paper, we discuss and quantify the normalized root-mean-squared (rms) transverse emittance $\epsilon_n = \sqrt{ \left\langle x^2 \right\rangle \left\langle p_x^2\right\rangle - \left\langle xp_x\right\rangle^2} / mc$, where $\left\langle \cdot \right\rangle$ represents averaging over the beam distribution, $m$ the electron mass and $c$ the speed of light. We also adopt the convention that the drive laser propagated in the $\hat{z}$ direction where the other direction in the simulation plane is the $\hat{x}$ direction. In some cases, the laser is polarized in the simulation plane (along $\hat{x}$) and in other cases not in the simulation plane (along $\hat{y}$). We label the emittance in the $x-p_x$ plane from a simulation with a subscript $x$ and $y$ depending on the polarizing direction. In ionization injection, $\epsilon_n$ is primarily determined by the initial state of the electrons at the instant of ionization and their immediate motion in the nonlinear wake and the laser. The electrons that become trapped, i.e., the K-shell electrons, are ionized within a volume where the electric field of the laser pulse is large enough to induce tunneling ionization \cite{ADKionizationmodel1986}. The size of this volume depends on both the intensity profile and the spot size of the laser, and it is typically a small fraction of the laser spot size. In the simulation shown in Fig. \ref{fig: simulation A}, the rms size of this column $\sigma_{x0}$ is 2.7 $\micro\meter$, about one fifth of the laser spot size $w_0$. The ionized electrons also acquire a transverse momentum (even if they are assumed to be born at rest) which along with the initial spot size defines the initial emittance or so-called thermal emittance. This initial momentum comes from two physical effects, namely the 1D residual drift and the transverse ponderomotive acceleration \cite{PhysRevLett.82.1688}. The 1D residual drift can be readily deduced from the conservation of the transverse canonical momentum of a charged particle which is only strictly held in a 1D plane wave \cite{kaw1973relativistic}. Even for finite width waves this gives a reasonable estimate for typical spot sizes. The normalized momentum of this drift is equal in magnitude to the normalized vector potential when and where the particle is ionized. Due to the fact that most electrons are released near the peak of the laser electric field by tunneling, where the vector potential is close to zero, this residual drift points in the polarization direction and it is therefore smaller than the simple estimation based on the peak value of the vector potential $a_0$ \cite{PhysRevLett.62.1259}. It can be also shown that for a linearly polarized laser polarized along $\hat{x}$, the electric field scales as $\epsilon a_0$ in the $\hat{z}$ direction and as $\epsilon^2 a_0$ in the $\hat{y}$ direction where $\epsilon=1/k_0w_0$ \cite{PhysRevE.58.3719}. Therefore, for linearly polarized laser the residual drift from this effect is very asymmetric between the $\hat{x}$ and $\hat{y}$ directions.

The contribution of the pondermotive force to the initial momentum can be estimated as $p_x \sim \left(a_0^2 c \tau_{FWHM} \right) / \left(\bar{\gamma} w_0\right) mc$, where $\bar{\gamma}$ is the average relativistic factor of the electron when it moves in the laser and $\tau_{FWHM}$ is the full-width-half-maximum (FWHM) duration of the laser pulse intensity. For cigar shaped pulses where $a_0^2 c \tau_{FWHM} \gg w_0$ or relativistically intense pulses $a_0 \gtrsim 1$ this may be the dominant cause for a residual drift. This leads to drifts in both directions perpendicular to the laser propagation direction. Furthermore, even for finite width waves there is a constant of the motion given by $\gamma-v_\phi p_z = 1+ \psi$ where $\gamma \equiv \sqrt{1+p^2/m^2c^2}, \psi \equiv e (\phi - v_\phi A_z)$ is the pseudo potential, $\phi$ is the scalar potential, $A_z$ is the axial vector potential, and $v_\phi$ is the phase velocity of the wake. This constant of the motion shows that even in the absence as a wake, i.e., $\psi = 0$, if a particle has a drift along the transverse direction it also has a drift along the laser propagating direction, i.e., $p_z/mc \approx \left( p_\perp/mc \right) ^2/2$.

In the simulation shown in Fig. \ref{fig: simulation A}, the residual transverse momenta in the $\hat{x}$ direction are $\sigma_{p_{0,x}}=0.65~mc$ and $ \sigma_{p_{0,y}}=0.41~mc$ respectively, where $x$ and $y$ refer to cases where the laser is polarized in ($\hat{x}$) or out ($\hat{y}$) of the simulation plane. The momenta for the $\hat{x}$ case is from both the 1D residual drift and the ponderomotive force while in the $\hat{y}$ case it is only from the ponderomotive force. Therefore, for this example both effects are important, but the ponderomotive acceleration is more important.

\begin{figure}[bp]
\includegraphics[width=0.5\textwidth]{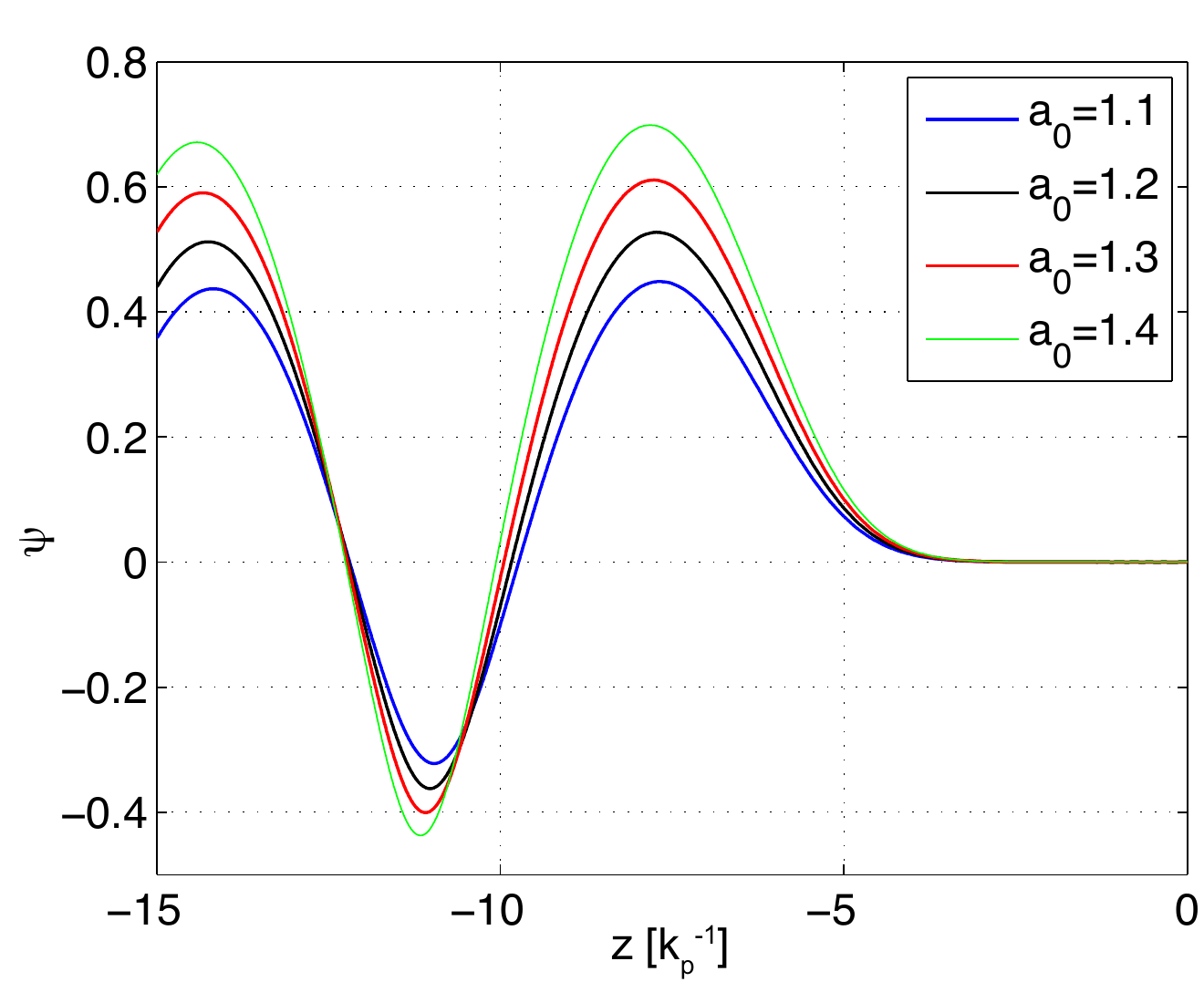}
\caption{\label{fig: psi v.s. a0} The normalized pseudo potential $\psi$ of the plasma wake on axis for different $a_0$ while keeping the intensity profile of the laser pulse fixed. }
\end{figure}

A simple and intuitive estimate of the beam emittance is the product of the initial transverse size $\sigma_0$ and the residual momentum $\sigma_{p_{x0}}$, $\epsilon_{th} = \sigma_{x_0}\sigma_{p_{x0}}$. We define this as the thermal emittance of the ionization-injection electrons. However due to the transverse betatron oscillation induced by the focusing force within the wake \cite{PhysRevLett.96.165002}\cite{lu2006nonlinearPoP}, the real beam emittance undergoes a complicated evolution process, as explained in detail in Ref. \cite{PhysRevLett.112.035003}. The emittance evolves from the thermal emittance because the electrons ionized at different times rotate in the transverse phase space at different betatron phases, leading to phase mixing, emittance growth and saturation unless the injection time is much shorter than $\omega_p^{-1}$. In Fig. \ref{fig: simulation A}(c), we plot the emittance evolution from simulation A in Table \ref{tab: simulation paras}. The initial growth and saturation of the beam emittance in the $x-p_x$ plane can be clearly observed. The emittance eventually saturates at a value approximately given by $\epsilon_{sat} \propto \left(k_p\sigma_{x_0}\right)^2/2 + \left( \sigma_{p_{x0}}/mc \right)^2$ \cite{Xu2013prst1}. For ionization injection by a single laser pulse, the typical value of emittance is about $1\micro\meter$ along the laser polarization direction (the laser was polarized in $\hat{x}$ direction) and few hundreds $\nano\meter$ perpendicular to the laser polarization (the laser was polarized in the $\hat{y}$ direction). One can see that both the thermal emittance $\epsilon_{th}$ and the saturated emittance $\epsilon_{sat}$ are determined by the values of $\sigma_{x_0}$ and $\sigma_{p_{x0}}$. Therefore, developing concepts to reduce both of these values could lead to lower final emittance.

\begin{figure}[bp]
\includegraphics[width=1.0\textwidth]{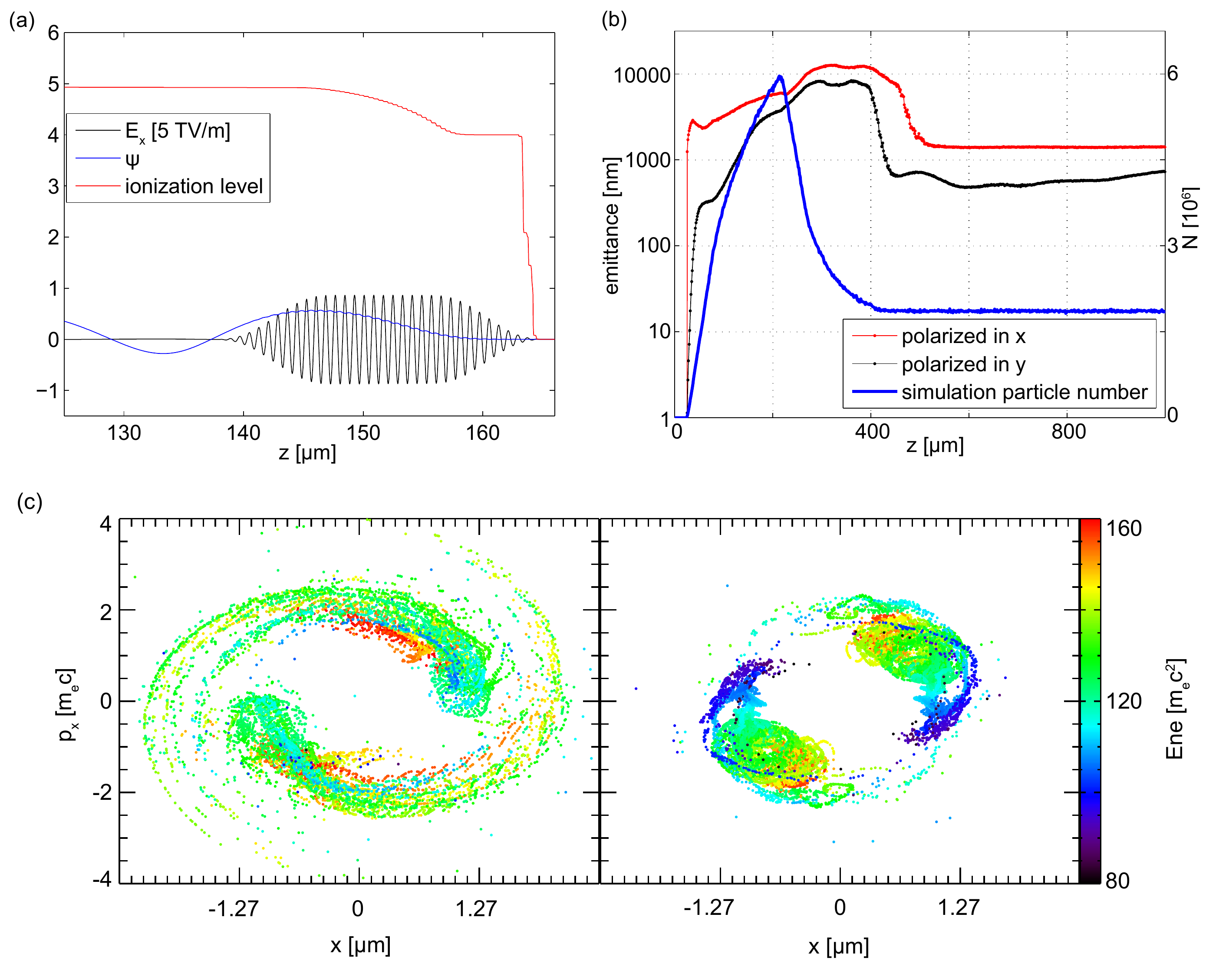}
\caption{\label{fig: simulation B} 2D OSIRIS simulation (B) of a 800 $\nano\meter$ laser pulse with $a_0=1.4$ propagates through a mix of pre-ionized plasma ($n_e=2.1\times 10^{18}~\centi\meter^{-3}$) and $C$ ($n_C = 0.01 n_e$). (a) The laser electric field $E_x$, the normalized pseudo potential $\psi$ and the ionization state of carbon atoms on axis. The laser longitudinal profile is a gaussian like 5th order symmetric polynomial $10t'^3 - 15 t'^4 + 6t'^5$, where $t'=(t-t_0)/t_{rise}$ for the envelope rise and $t'=(t_0 + t_{flat} - t)/t_{fall}$ for the envelope fall. In this simulation $t_{rise}=t_{fall}=34~\femto\second, t_{flat}=25~\femto\second$. (b)The emittance evolution along (red) and perpendicular to the polarization direction (black), and the evolution of the particle number in the simulation (blue). The injected charge is about 20 $\pico\coulomb$ by assuming a cylindrical distribution around $z$-axis. (c) The $x-p_x$ phase spaces at $z=1~\milli\meter$ for the laser polarized in (left) and out (right) of the simulation plane.}
\end{figure}

For the single laser pulse scheme of ionization injection, one obvious possibility for reducing $\sigma_{x_0}$ and $\sigma_{p_{x0}}$ is to decrease the laser intensity. However the trapping condition \cite{PhysRevLett.104.025003} for electron injection $\delta \psi \approx -1$ places a lower limit on the wake amplitude and hence a lower limit on the laser intensity. The electrons can be trapped if $\psi_{min} - \psi_{birth} < -1$, where $\psi_{min}$ is the minimum value of the pseudo potential and $\psi_{birth}$ (electrons can be born off axis) is the pseudo potential where the electrons are released. Fig. \ref{fig: psi v.s. a0} shows the pseudo potentials for different driver intensities. The normalized vector potential of the laser $a_0$ is varied while the laser intensity profile was kept the same as in the simulation A. It is found that the minimum $a_0$ for $\left( \delta\psi \right)_{min}\approx -1.0$ is about $1.3$. We note that to use this intensity to reach injection, carbon needs to be chosen instead of nitrogen to provide the K-shell electrons for trapping due to their relatively lower IP than that of nitrogen and the ionization needs to occur at the maximum of $\psi$. In Fig. \ref{fig: simulation B}, a 2D PIC simulation (simulation B) for $a_0=1.4$ is shown to confirm that the injection does occur at this low intensity. However the emittances obtained in this case, $\sim1500~\nano\meter (\hat{x})$ and $\sim500~\nano\meter (\hat{y})$, are not better than those obtained in the simulation A. This indicates that it is difficult to further reduce the emittance by just fine tuning the laser parameters of the single laser pulse for ionization injection. In this simulation the $\left( \delta\psi \right)_{min}$ is close to -1, therefore only electrons with large initial transverse positions which lead to large transverse momenta when they cross the $z$-axis are injected ($\delta\psi < -1 + \sqrt{1 + p_\perp^2/m^2c^2} / \gamma_\phi$) \cite{PhysRevLett.104.025003}\cite{lu2006nonlinear} and form a ring structure in 2D simulations. The $x-p_x$ phase space plots are shown in Fig. \ref{fig: simulation B}(c).

\section{LWFA with ionization injection using two laser pulses with different wavelength --- longitudinal injection}
As mentioned earlier, in order to generate electron beams with emittances below 100 $\nano\meter$, the transverse size of the electrons at birth and their residual momentum need to be further reduced. Intuitively, a second injection laser pulse with the same wavelength and much smaller spot size may appear to be the solution. However, simulations and analytical estimates show that for such a scheme the electrons will a large transverse residual momenta due to the ponderomotive acceleration ($\propto \nabla I$) induced by the small spot size of the injection laser. Here we describe results from simulation C where a circularly polarized 800 $\nano\meter$ drive laser pulse with an $a_0=1.2$, which is high enough to drive a nonlinear wake but low enough as not to ionize the K-shell electrons of nitrogen, is propagating through a gas mixture. An injection pulse has the same wavelength, a focal spot of 2.5 $\micro\meter$, a pulse duration of 10 $\femto\second$ and an $a_0=2.0$, which is intense enough to ionize the K-shell electrons of nitrogen. With these parameters, the spot size of the injected electrons is reduced to $\sigma_{x0}= 0.71~\micro\meter$, but the residual momenta, $\sigma_{p0, x} =0.77~mc, \sigma_{p0, y}=0.58~mc$, are still quite high due to the ponderomotive acceleration, which leads to final emittances $\sim1600~\nano\meter (\hat{x})$ and $\sim800~\nano\meter(\hat{y})$.

\begin{figure}[bp]
\includegraphics[width=1.0\textwidth]{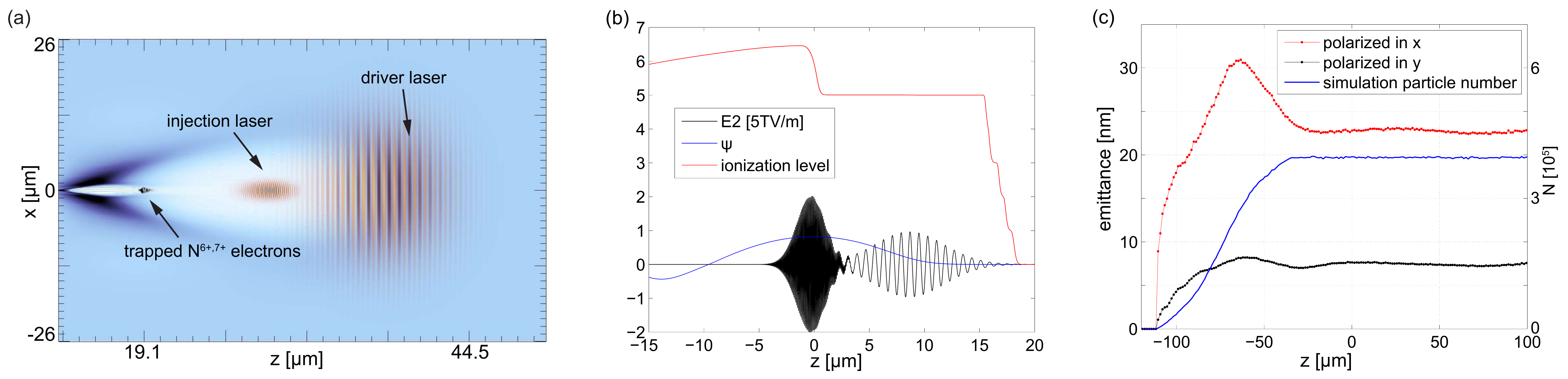}
\caption{\label{fig: simulation D} 2D OSIRIS simulation (D) of the two color ionization injection scheme using a 800 $\nano\meter$ laser as the drive laser and a 80 $\nano\meter$ laser for injection ($n_e=1.7\times 10^{18}~\centi\meter^{-3}$). The  circularly polarized 800 $\nano\meter$ laser pulse with spot size 10 $\micro\meter$ and $a_0=1.2$ propagates to the right in a mixture of $\mathrm{N}^{5+}$ ($n_{\mathrm{N}^{5+}} = 0.2~n_e$) and pre-ionized plasma. The  linearly polarized 80 $\nano\meter$ laser pulse with spot size 0.64 $\micro\meter$ and $a_0=0.25$ is used to ionize and inject the electrons into the wake. The longitudinal profile of the driver laser and the injection laser are both gaussian with $\tau_{FWHM}=18~\femto\second$ and $7~\femto\second$, respectively. $z=0~\micro\meter$ is the focal plane of the injection laser. (a) Snapshot of  the charge density distribution of the wake electrons, the  K-shell electrons of nitrogen, and the electric field in $x$ direction. (b)  The emittance evolution along (red) and perpendicular to the polarization direction (black), and the evolution of the particle number in the simulation (blue). The injected charge is about 1.5 $\pico\coulomb$ by assuming a cylindrical distribution around $z$-axis.}
\end{figure}

To reduce the residual momenta of the injected electrons significantly, an injection laser with much shorter wavelength is needed, we call this the two color scheme. For such an injection laser, the vector potential $a_0$ that is needed for freeing the electrons through tunneling ionization is much reduced (due to the simple relation $a_0 \propto E_0 \lambda$), therefore it can significantly reduce the residual momenta. As we noted in the introduction, this is inherently a multi-dimensional process. Next, we give an example from 2D PIC simulation by using an injection laser with $\lambda=80~\nano\meter$ (simulation D). Additional detail is given in Table \ref{tab: simulation paras} and in the figure capture to Fig. \ref{fig: simulation D}. As shown in Fig. \ref{fig: simulation D}(a), the injection laser with $a_0=0.25$ is focused down to 0.64 $\micro\meter$ to ionize the K-shell electrons of nitrogen in a very small transverse region. The transverse birth size is $\sigma_{x0}=0.24~\micro\meter$, and the residual momenta for the laser polarized in and out the simulation plane are $\sigma_{p0, x}=0.066~mc, \sigma_{p0, y}=0.023~mc$ respectively. The injection distance is about 77 $\micro\meter \approx 19 c/\omega_p$, which is long enough for transverse phase mixing and emittance saturation to occur \cite{PhysRevLett.112.035003}. As shown in Fig. \ref{fig: simulation D}(b), the emittances saturate at 20 $\nano\meter (\hat{x})$ and 8 $\nano\meter (\hat{y})$, where () refers to the polarization direction of the laser, which are significantly better than for the single wavelength case.

\begin{figure}[bp]
\includegraphics[width=1.0\textwidth]{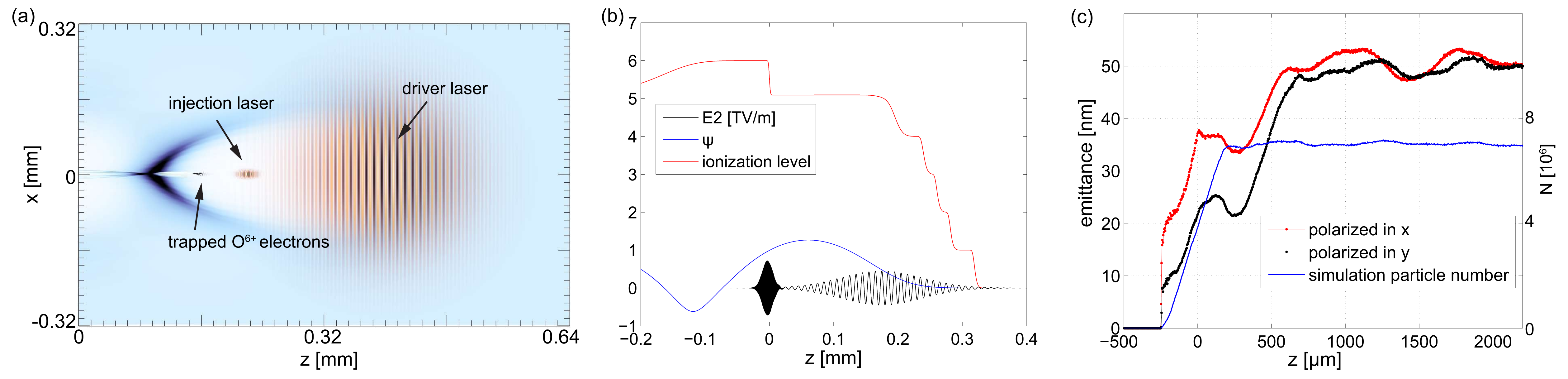}
\caption{\label{fig: simulation E} (a) 2D OSIRIS simulation (E) of the two color ionization injection scheme using a 10 $\micro\meter$ driving laser and a 400 $\nano\meter$ injection laser. The density of the pre-ionized plasma electrons and the $\mathrm{O}^{5+}$  ions are $n_e=1.19\times 10^{16}~\centi\meter^{-3} $ and $n_{\mathrm{O}^{5+}}=0.2n_e$ respectively. The driving laser has a longitudinal gaussian profile with $\tau_{FWHM}=~352 \femto\second$, and with a transverse spot size $w_0= 142~\micro\meter$.  The injection laser also has  a longitudinal gaussian profile with $\tau_{FWHM}=44~\femto\second$, and with a transverse spot size  $w_1= 4.8~\micro\meter$. $z=0~\milli\meter$ is the focal plane of the injection laser. (b) The laser electric field $E_x$, the normalized pseudo potential $\psi$ of the plasma wake and the ionization state of oxygen atoms on axis. (c)The emittance evolution along (red) and perpendicular to the polarization direction (black), and the evolution of the particle number in the simulation (blue).}
\end{figure}

In the previous simulation, the laser power is about 85 $\giga\watt$ and the total energy within the pulse is about 0.6 $\milli\joule$ which is sufficient to ionize the sixth electron of N. Currently such a powerful laser at 80 $\nano\meter$ does not exist. For example extreme ultraviolet (EUV) sources based on high-order harmonics generation (HHG) typically have energies of a few  $\micro\joule$ \cite{lambert2005seeding}. Therefore an experiment based on the above parameters will not be possible without improvement in the performance of such sources. 

We therefore examine in 2D PIC simulations using a 10 $\micro\meter$ laser pulse to drive the wake and a 400 $\nano\meter$ laser pulse to inject electrons into this wake. Lasers with a longer wavelength have a larger vector potential for fixed laser intensity, therefore it is advantageous to use such a longer wavelength laser with large ponderomotive potential to drive nonlinear wakes and lower intensity to limit K-shell ionization. High power short pulse lasers needed for the injection pulse (with energy more than 10 $\milli\joule$) can be obtained for wavelengths as short as 200 $\nano\meter$ by frequency up-conversion of a 800 $\nano\meter$ lasers. On the other hand, high power lasers with a mid-infrared wavelength to generate the wake are currently under development. For example, a 100 $\tera\watt$ short pulse CO$_2$ (10 $\micro\meter$ wavelength ) laser is currently under construction at Accelerator Test Facility of Brookhaven National Laboratory \cite{polyanskiy2013ultrashort} while a 15 $\tera\watt$  peak power CO$_2$ laser is currently operational, albeit with a 2 $\pico\second$ laser pulse at UCLA \cite{haberberger2010fifteen}. Such a scheme is therefore realizable experimentally with modest further improvements to existing technology.

In our simulation (simulation E) , oxygen is used for the injection gas, where the first five electrons from oxygen are used to form the wake, and the sixth electron of oxygen atoms is used to supply the trapped electrons. We also use a preformed plasma. A circularly polarized laser ($\lambda_0 = 10~\micro\meter,a_0=1.4,E_{peak}=0.45~\tera\volt\per\meter$) drives a nonlinear wake in the blowout regime. A linearly polarized short wavelength laser ($\lambda_1 = 400~\nano\meter,a_1=0.09,E_{peak}=0.72~\tera\volt\per\meter$) trailing the driving laser is tightly focused to ionize the sixth electron with small transverse size and low residual momentum. In Fig. \ref{fig: simulation E}(a) 2D contour plot of the drive laser, injection laser, background electrons and trapped O$^{5+}$ electrons are shown. Lineouts of the lasers, the wake potential and the ionization levels of the oxygen are shown in Fig. \ref{fig: simulation D}(b). The power of the driver and the injection laser are 17 $\tera\watt$ (6.4 $\joule$ in 352 $\femto\second$) and 22 $\giga\watt$ (1 $\milli\joule$ in 44 $\femto\second$), respectively. 

\begin{figure}[bp]
\includegraphics[width=1.0\textwidth]{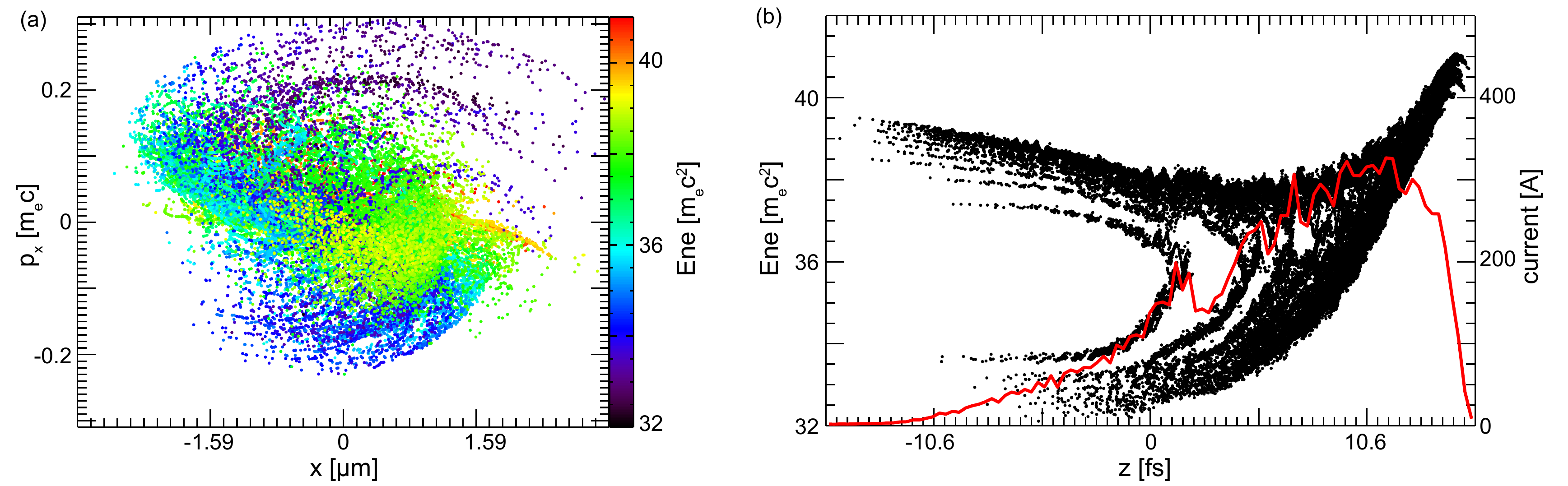}
\caption{\label{fig: phase space E} Snapshots of the phase space from simulation (E) of Fig.5. (a) the transverse phase space at $z=2.2~\milli\meter$; (b) the longitudinal phase space and the current profile  at $z=2.2~\milli\meter$.}
\end{figure}

The birth size and the residual momentum of the trapped electrons are $\sigma_{x0} =1.2~\micro\meter, \sigma_{p0, x} = 1.8\times 10^{-2}~mc, \sigma_{p0, y} = 2.3\times 10^{-3}~mc$, respectively. The emittance evolution of the injected electron beam  is shown in Fig. \ref{fig: simulation E}(c), and one can see that the emittances saturate at 50 $\nano\meter$ after the initial fast growth. The total injected charge is estimated as 4 $\pico\coulomb$ by assuming the injected beam is cylindrically symmetric along $z$-axis. We note that due to the relatively long injection distance ($0.45~\milli\meter \approx 9c/\omega_p$), the slice energy spread of the injected beam reaches a few \mega\electronvolt,  as shown in Fig. \ref{fig: phase space E}(b).

\section{LWFA with ionization injection using two laser pulses with different wavelengths --- transverse injection}
\begin{figure}[bp]
\includegraphics[width=1.0\textwidth]{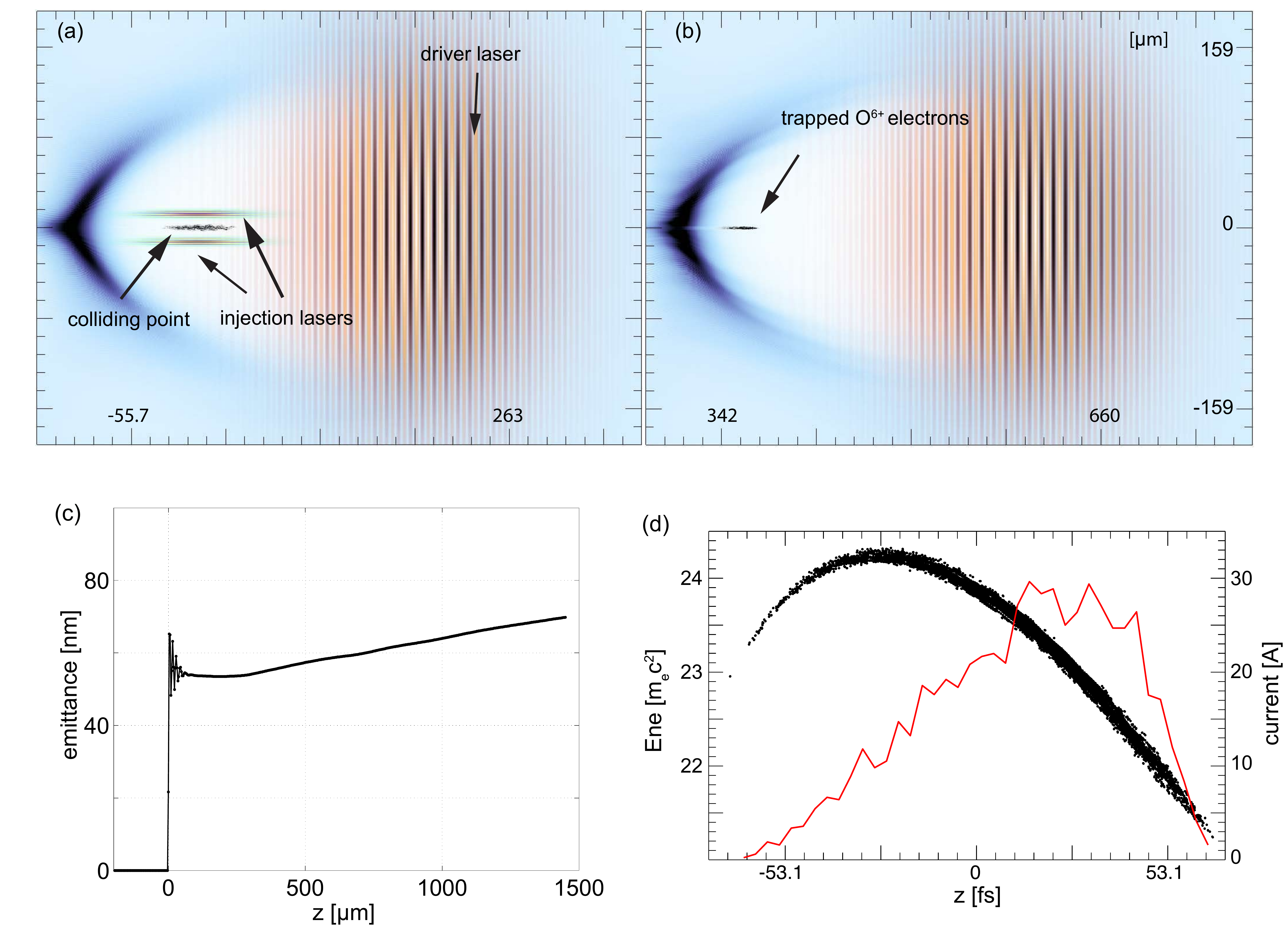}
\caption{\label{fig: simulation F} OSIRIS simulations of the two color ionization injection scheme by two transverse colliding pulses ($n_e=1.12\times 10^{16}~\centi\meter^{-3}$). The  circularly polarized 10 $\micro\meter$ laser pulse with spot size 142 $\micro\meter$, $\tau_{FWHM}=352~\femto\second$ and $a_0=1.4$ propagates to the right in a mixture of $\mathrm{O}^{5+}$ ($n_{\mathrm{O}^{5+}} = 0.2~n_e$) and a pre-ionized plasma. The two linearly polarized 400 $\nano\meter$ laser pulse with spot size 48 $\micro\meter$ and $a_0=0.06$ ionize the $\mathrm{O}^{5+}$ to generate the trapped electrons. $z=0~\milli\meter$ is the plane where the two injection laser pulses --- spatially and temporally overlap.The two injection laser pulses have a gaussian longitudinal profile with $\tau_{FWHM}=13~\femto\second$. (a) Snapshot of the charge density distribution of wake electrons, the K-shell electrons of nitrogen, and the electric field in $x$ direction and in $z$ direction at the laser pulses's collision time, (b) 1 $\pico\second$ after the collision when the injected electrons become trapped in the wake. (c) The emittance evolution in $x$ direction. The injected charge is about 2 $\pico\coulomb$ by assuming a cylindrical distribution along $x$-axis. (d) The $z-p_z$ phase space and the current profile of the injected beam at $z=1.5~\micro\meter$.}
\end{figure}

In the above longitudinal injection scheme, the intensity of the injection laser is large enough to release the trapped electrons when the pulse is near its focal plane. Before and after the focal plane, the intensity of the laser pulse decreases and the ionization ceases. So the injection distance is on the order of the Rayleigh length of the injection laser which is about several or dozens of plasma skin depths for typical parameters. The energy spread in each slice of the injected beam grows with the injection distance, and typically reaches a few MeV by the end of the injection process. To significantly reduce the slice energy spread, we propose  to use two transverse colliding pulses to limit the injection distance \cite{PhysRevLett.111.015003}. This has also been shown to reduce the emittance to a beam driver. Two counter-propogating short wavelength laser pulses ($\lambda=400\nano\meter$) moving along the + and - $x$-axis directions are synchronized with the driver laser so that they overlap inside the nonlinear wake near the point where the longitudinal electric field vanishes. In Ref. \cite{PhysRevLett.111.015003} where a particle beam driver was used the ionizing lasers had an $a_0$ of $0.016$ as wavelength of 800 $\nano\meter$, and the ionizing gas was He. Because we are using a laser driver which has a larger intensity, oxygen is used and the ionizing lasers have a normalized potential $a_0=0.06$ (not ionize the sixth K-shell electron of oxygen). The laser intensity exceeds the ionization threshold only where the two lasers overlap, and a large fraction of $\mathrm{O}^{5+}$ ions within this volume is now ionized to $\mathrm{O}^{6+}$ as shown in Fig. \ref{fig: simulation F}(a). As the lasers travel past the collision point, the injection ceases. These laser-ionized oxygen electrons then respond to the wake fields and are rapidly accelerated to a longitudinal velocity close to c as they slip backwards to the rear of the nonlinear wake. They then began to move nearly synchronously with the wake, as depicted in Fig. \ref{fig: simulation F}(b). Fig. \ref{fig: simulation F}(c) shows the emittance evolution in this simulation, and one can see that emittance of $\sim60~\nano\meter$ is achieved. Fig. \ref{fig: simulation F}(d) shows the longitudinal phase space of the injected beam, which confirms clearly that the slice energy spread (as small as 30 $\kilo\electronvolt$) is much reduced comparing with the longitudinal injection case (a few $\mega\electronvolt$). We note that due to the fact that $a_0=0.06$ and $\lambda=400~\nano\meter$ rather than $0.016$ and $\lambda=800~\nano\meter$that the emittance is larger than for a particle beam driver. The released electrons conduct oscillations in the laser propagation direction under the electric field of the standing wave. The residual momentum in the propagation direction is determined by the normalize vector potentials and the period of the laser pulses. In our simulation, a larger laser intensity for the ionizing law leads to a larger residual momentum than the value in the particle beam driver case. Understanding these tradeoffs is an area for future study.

\section{Conclusion}
In this paper we have examined ionization injection into plasma wave wakes, where both the wake and the injection are driven by lasers. We carried multi-dimensional PIC simulations using OSIRIS. As a reference case, we simulated ionization injection from a single laser. We then studied improvements from using a separate laser (with the same wavelength) with a tighter focus to trigger the ionization. Finally, we studied the improvement in beam quality that results when the drive and ionizing lasers have different wavelengths (colors), including having the incoming lasers co-propagate or transversely propagate with respect to the drive laser. For a fixed laser intensity, lasers with longer(shorter) wavelengths have larger(smaller) ponderomotive potentials ($a_0\propto E_0 \lambda$). The two-color scheme utilizes this property to separate the injection process from the wakefield excitation process. Very strong wakes can be generated at relatively low laser intensities by using a longer wavelength laser driver (e.g. 10 $\micro\meter$ CO$_2$ laser) due to its relatively large ponderomotive potential. On the other hand, a short wavelength laser can produce electrons with very small residual momenta ($p_\perp\sim a_0$) inside the wake, leading to electron beams with reduced normalized emittances (tens of $\nano\meter$). We find that very bright electron beams can be generated through this two-color scheme in either collinear propagating or transverse colliding geometry. Our 2D particle-in-cell simulations indicate that a 10 $\femto\second$ electron bunch with $\sim4~\pico\coulomb$ of charge and a normalized emittance of $\sim 50~\nano\meter$ can be generated by combining a 10 $\micro\meter $ driving laser with a 400 $\nano\meter$ injection laser, which is an improvement of more than one order of magnitude compared to the typical results obtained when a single wavelength laser is used for both the wake formation and ionization injection. Such an experiment should be feasible in the near future.

Work supported by NSFC Grants No. 11175102, No. 11005063, Tsinghua University Initiative Scientific Research Program, the Thousand Young Talents Program, DOE grants DE-FG02-92-ER40727, DE-SC0008491, DE-SC0008316, and NSF Grants No. PHY-0936266, No. PHY-0960344, and ACI-1339893. Simulations are performed on Hoffman and Dawson2 clusters at UCLA and NERSC at LBNL.

\bibliography{refs_xinlu}

\begin{thebibliography}{35}
\bibitem [{Xu2()}]{Xu2013prst1}%
  \BibitemOpen
  \href@noop {} {}\bibinfo {note} {The paper will be presented in the
  future.}\BibitemShut {Stop}%
\end{thebibliography}

\end{document}